\documentclass[showpacs,12pt]{revtex4}

\usepackage{graphicx}%
\usepackage{dcolumn}
\usepackage{amsmath}

\begin{document}

\preprint{HEP/123-qed}

\title{\bf Fluctuation Studies at the Subnuclear Level of
Matter: \\Evidence for Stability, Stationarity and Scaling \footnote{Supported
by the National Natural Science Foundation of China under Grant No.70271064.}}

\author{LIU Qin$^{1}$\footnote{Email address: liuq@ccnu.edu.cn}}
\author{ MENG Ta-chung$^{1,2}$\footnote{Email address: meng@ccnu.edu.cn; meng@physik.fu-berlin.de}}%
\affiliation{\it $^{1}$Department of Physics, CCNU, 430079 Wuhan, China
\\ $^{2}$Institut f\"{u}r Theoretische Physik, FU-Berlin, 14195 Berlin, Germany}


\date{\today}

\begin{abstract}
It is pointed out that the concepts and methods introduced by Bachelier and
 by Mandelbrot to Finance and Economics can be used to examine the
fluctuations observed in high-energy hadron production processes.
Theoretical arguments and experimental evidences are presented which
show that the relative variations of hadron-numbers between successive
rapidity intervals are non-Gaussian stable random variables, which exhibit
stationarity and scaling. The implications of the obtained results are discussed.
\end{abstract}


\maketitle


\section{Hadron-production in high-energy collisions}

It is a well-known fact that hadrons (for example pions, kaons,
protons, neutrons, $\cdots$ , and their antiparticles) can be produced
in lepton-, hadron- and nucleus- induced reactions at sufficiently
high incident energies. In such processes, we are witnessing an impressive
demonstration of energy-conversion into matter, where energy-momentum
 conservation requires that the total (c.m.s.) energy of the colliding
 system must be high enough to creat the masses of the produced hadrons
 and provide them with sufficient kinetic energies. A vast amount
 of data exists for such hadron-formation processes where the
 multiplicity-, transverse-momentum- and rapidity-distributions of
 the produced hadrons in high-energy hadron-hadron, hadron-nucleus and
 nucleus-nucleus collisions have been accumulated, and part of them, especially
 those at extremely high energies are taken from cosmic-ray experiments.

 Much effort has been made to describe these data. The conventional
 way of doing this is to divide such a hadron-production
 process, in accordance with the currently most popular picture
 for subnuclear structure and subnuclear dynamics, conceptually into three steps:

 In step I, the incident-hadron (a free hadron or a bounded hadron namely
 that inside a nucleus) is pictured as a large swarm of partons
 (quarks, antiquarks and gluons) where everyone of them carries a
 fraction (known as the Feynman-$x$ and denoted by $x_F$ or simply by $x$,
 with $x_{F}$ or $x$ between zero and unity) of the
 longitudinal (that is along the incident axis) momentum of this hadron.
 The distribution of $x_F$ for different kinds of partons (for example: the u,
 d, s valence quarks, the sea-quark-pairs; or the gluons) has been
 carefully parameterized; and the corresponding parton distributions
 can be readily found in the literature \cite{1}.

 In step II, it is envisaged that some of the partons in the projectile-hadron are
 scattered by some of the partons in the target-hadron. They
 interact with one another according to the Feynman rules derived from
 the given QCD-Lagrangian. Under the assumption that some classes of
 reaction-mechanisms are more important than the others and
 the perturbative methods are applicable, one can evaluate the
 corresponding Feynman graphs. Details of such calculations have
 been worked out, and can be readily found in the literature \cite{1}.

 In step III, the observed final-state hadrons are assumed to
 be directly related to the scattered partons mentioned in Step II.
 The relations between the scattered partons and the observed
 hadrons are parameterized in terms of ``fragmentation functions''.
 Fragmentation functions for all possible kinds of partons have been
 worked out, and they can also be readily found in the literature \cite{1}.

 In connection with the above-mentioned three-step-approach, it is of considerable
 importance to recall \cite{1} the following facts: (A) Quarks, antiquarks and
 gluons have not been, and according to QCD and Confinement they can
 never be, directedly measured. (B) Perturbative
 methods can be used only when the momentum transfer in the scattering
 process is so large that the corresponding QCD running coupling constant is
 less than unity; but the overwhelming majority of such hadron-production
 processes are soft in the sense that the momentum transfer is relatively low.
 These facts imply that, it {\it has not been}
 possible, and it {\it can never be} possible,
 to have an {\it experimental check} for the
 three steps {\it individually}.

 In the present paper, we discuss such hadron-production processes
 from a different viewpoint and ask the following questions: Suppose
 we focus our attention {\it only} on the directly measurable
 quantities (such as the rapidity distributions of the produced
 charged hadrons), and try to perform a preconception-free
 data-analysis where we rely as much as possible on the relevant
 knowledge of Mathematical Statistics [such as the Law of Large
 Numbers and the (Generalized) Central Limit Theorem], shall we be able to extract
 useful information on the reaction mechanisms of such hadron-production
 processes by examining the experimental data? If yes, what can we
 learn from such information? Can such information be used, for example,
 to make predictions for future experiments, and/or to check the
 consequences of the conventional approach mentioned above?

 Since a considerable part of the available first-hand information on
 hadron-production processes are distributions of few
 directly measurable quantities (such as the multiplicity $n_{ch}$ of
 charged hadrons; the rapidity $y$ or the pseudorapidity $\eta$ and the
 transverse momentum $p_{\perp}$ or transverse energy $E_{\perp}$
 of such a hadron), concepts and methods in Probability Theory and Mathematical
 Statistics are expected to play a distinguished role in describing/understanding
 the existing data, and in making predictions for future experiments.

\section{From random walk to stable distributions in Finance and Economics}
It is well-known that fluctuation studies are of considerable
importance in non-equilibrium as well as in equilibrium systems;
and there are several reasons for this. One of them is that,
deviations for or fluctuations about the mean values are always
present, even when the system under consideration is in equilibrium.
Another is that fluctuation studies provide a natural framework for
understanding a large classes of phenomena, among which the best
known phenomenon is ``Brownian motion'' or ``the random walk''---known
through Albert Einstein's work starting 1905 \cite{2}.

It is, however, not very well-known (at least not among the
particle physicists) that, five years earlier, in 1900, a then
young French student named Louis Bachelier has constructed a
random walk theory for security and commodity markets \cite{3}.
The essence of Bachelier's theory is the following: Consider the
successive differences of two adjacent spot prices
\begin{equation}
L_{B}(t,T)=z(t+T)-z(t),
\end{equation}
where $z(t+T)$ and $z(t)$ is the spot price of a stock or that of
a piece of commodity at the end of a period of time $t+T$ and $t$
respectively; and assume that the set of all $L_{B}(t,T)$'s are
statistically independent, Gaussian (normally) distributed random
variables with zero mean, and variance proportional to the
differencing interval $T$. This assumption is often called
\cite{4} the Gaussian hypothesis \cite{3}, and the arguments
supporting this hypothesis are based on the Central Limit Theorem
(cf. Appendix A and the references given therein). To be more
precise, if the price changes from transaction to transaction are
independent, identically distributed random variables with {\it
finite} variance, and if transactions are {\it fairly uniformly}
spaced through time, the Central Limit Theorem will lead us to
believe that price changes across differencing intervals such as a
day, a week, or a month should be Gaussian distributed, because
they are simple sums of the changes from transaction to
transaction. Detailed comparisons between this hypothesis and
empirical data have been performed by a number of authors
\cite{4}. Despite the fundamental importance of Bachelier's work
\cite{3}, it has been become obvious that this hypothesis cannot
be right as it stands (some details will be given in Section IV).
In this sense, Mandelbrot's theory \cite{5,6,4} can be understood
as an improvement and a generalization of Bachelier's \cite{3}.

The main difference between the theory of Bachelier and that of
Mandelbrot is the following. Instead of Gaussian (normally
distributed) random variables with zero mean and finite (which can
be normalized to unity) variance, Mandelbrot made the assertion
that the variance of the random variable
\begin{equation}
L_{M}(t,T)=\ln z(t+T)-\ln z(t)
\end{equation}
can behave as if it is {\it infinite}; and consistent with this,
he suggested that the Gaussian distribution should be replaced by
a rich class of probability distributions called {\it stable
distributions} among which Gaussian is a limiting case with finite
variance. It is known (cf. Appendix A and the references given
therein) that such distributions allow skewness and heavy tails,
and have many intriguing mathematical properties. One of the
important properties is that, instead of the above-mentioned
Central Limit Theorem, they satisfy the {\it Generalized} Central
Limit Theorem (cf. Appendix A and the references given therein).
Above all, evidences have been found that the empirical
distributions conform best to the non-Gaussian members of this
family of distributions.

It is of considerable importance to keep in mind that the
statistical distributions in Mandelbrot's Fractal Approach to
Finance and Economics are {\it stable}, {\it stationary}, and {\it
scale-invariant} (in other words, {\it scaling}). Furthermore,
such distributions can have one or several of the following
properties. (a) Repeated instances of sharp discontinuity can
combine with continuity. (b) Concentration can automatically and
unavoidably replace evenness. (c) Non-periodic cycles can
automatically and unavoidably follow from long-range dependence.
While some of the mathematical aspects of the above-mentioned
concepts are summarized in Appendix A, the methods and results of
comparing such concepts with experimental data will be discussed
in Section IV.

\section{Rapidity, locality and light-cone variables in high-energy hadron-production processes}
Quantitative information about the produced hadrons in high-energy
collisions exists mostly in form of distributions (histograms) of
the measured quantities; and one of the few distributions that can
be, and has been, well measured is the rapidity ($y$) or
pseudorapidity ($\eta$) distribution of the electrically charged
hadrons produced in such processes.

The rapidity of an observed hadron of mass $m$, energy $E$,
momentum $\vec{p}=(p_{\parallel},\vec{p}_{\perp})$ where
$p_{\parallel}$ and $\vec{p}_{\perp}$ are the momentum in the
parallel and that in the transverse direction of the collision
axis respectively, is defined as
\begin{equation}
y=\frac {1}{2}\ln(\frac{E+p_{\parallel}}{E-p_{\parallel}}).
\end{equation}
This can also be written as $y=\ln[(E+p_{\parallel})/E_{\perp}]$,
where $E_{\perp}=M_{\perp}=(m^{2}+\left|\vec{p}_{\perp}\right|^{2})^{1/2}$ is known
as the ``transverse energy'' or ``transverse mass''. Due to the fact that the
overwhelming part of the produced hadrons are energetic pions (the mass of which is
negligibly small compared with the average value of $p_{\parallel}$), and that their
transverse momenta are exponentially distributed with relatively small
average $\left|\vec{p}_{\perp}\right|$, the corresponding pseudorapidity
\begin{equation}
\eta=\frac{1}{2}\ln(\frac{p+p_{\parallel}}{p-p_{\parallel}})
\end{equation}
is often used as a good approximation for the rapidity $y$ ($y\approx\eta$), where $E$ in Eq. (3)
is replaced by $p=({p_{\parallel}}^{2}+\left|\vec{p}_{\perp}\right|^{2})^{1/2}$.
One of the advantages of using $\eta$ instead of $y$ is that the former can
be expressed in terms of the scattering angle $\theta$ ($\eta=-\ln\tan\frac{\theta}{2}$),
and thus in order to determine $\eta$ experimentally, one only needs to measure $\theta$.
Furthermore, we can also write $y\approx\eta=\ln[(p+p_{\parallel})/\left|\vec{p}_{\perp}\right|]$ and
treat $\left|\vec{p}_{\perp}\right|$ approximately as a constant.

Significant fluctuations have been observed in
rapidity-distributions in hadron-production processes at
cosmic-ray \cite{7}, and at accelerator energies \cite{8,9}. The
question we raise here is whether such fluctuations can be
directly used to gain information on the reaction mechanisms of
the above-mentioned processes in general, and information on the
space-time properties of such reactions in particular. In order to
study such properties, it is useful to keep in mind that the
overwhelming part of the produced hadrons are energetic pions
which implies that their locations in space-time are mainly
concentrated near the light-cone. Hence it is not only meaningful,
but also useful to discuss the space-time properties of these
hadrons in terms of light-cone variables \cite{10} (and/or
quantities directly related to them)
\begin{equation}
 x_{\pm}=\frac{1}{\sqrt{2}}(t\pm x_{\parallel}),
\end{equation}
\begin{equation}
 p_{\pm}=\frac{1}{\sqrt{2}}(E\pm p_{\parallel}),
\end{equation}
where $(t,x_{\parallel},\vec{x}_{\perp})$ and $(E,p_{\parallel},\vec{p}_{\perp})$
are canonical conjugates to each other with the property
that their 4-product can be written as
\begin{equation}
  tE-x_{\parallel}p_{\parallel}-\vec{x}_{\perp}\cdot\vec{p}_{\perp}=
  x_{-}p_{+}+x_{+}p_{-}-\vec{x}_{\perp}\cdot\vec{p}_{\perp}.
\end{equation}

Here, we introduce in analogy to rapidity $y$ defined by Eq. (3),
a quantity
\begin{equation}
  l=\frac{1}{2}\ln(\frac{t-x_{\parallel}}{t+x_{\parallel}})
\end{equation}
which we call locality. This quantity can, again in analogy to
$y$, also be written as $l=\ln[(t-x_{\parallel})/s_{\perp}]$ where
$s_{\perp}=(t^{2}-x_{\parallel}^{2})^{1/2}$ stands for the
``transverse interval'' whereby
$s^{2}=t^{2}-x_{\parallel}^{2}-{\left|\vec{x}_{\perp}\right|}^{2}$
is the square of the space-time ``interval'' between the
world-point $x$ and the origin $(0,0,\vec{0})$. Furthermore, we
can also define a corresponding ``pseudolocality''
\begin{equation}
 \lambda=\frac{1}{2}\ln(\frac{r-x_{\parallel}}{r+x_{\parallel}}),
\end{equation}
where $r$ in $s^{2}=t^{2}-r^{2}$ stands for
$r=({x_{\parallel}^{2}}+{\left|\vec{x}_{\perp}\right|}^{2})^{1/2}$.
While the dynamics of a system can be described either in terms of
space-time variables or in terms of their canonical conjugates, it
is certainly of considerable interest to have a physical picture
in space-time.

Dealing with subnuclear phenomena, it is clear that we need to
take care of the commutation relations \cite{11} between
quantities which are canonical conjugates to each other. From
$[E,t]=i$ and $[p_{\parallel},x_{\parallel}]=-i$, we readily
obtain the corresponding commutation relations not only between
the light-cone variables $x_{\pm}$ and $p_{\pm}$, but also those
between $l$ and $y$ as well as those between $\lambda$ and $\eta$.
While the derivation and the explicit expressions of these
relations are not very interesting, one of the consequences should
be mentioned: By using a very general form of W. Heisenberg's
uncertainty principle developed in 1929 by H. P. Robertson
\cite{12}, the following can be obtained. If we are interested in
the simultaneous specification of two observables $A$ and $B$, the
corresponding operators of which obey a non-zero commutation
relation $[A,B]=iC$ ($i$ is included for convenience; for
$A=x_{\parallel}$ and $B=p_{\parallel}$ we have $C=\hbar$), and if
we consider a state $|\psi>$ which is normalized but otherwise
arbitrary (i.e. not necessarily eigenstate of either $A$ or $B$),
then, the root mean square derivations of $A$ and $B$ as the
square-roots of the corresponding quantities satisfy $\Delta A
\Delta B \geq
\frac{1}{2}\left|<C>\right|=\frac{1}{2}\left|<[A,B]>\right|$,
which is an exact and precise form of the uncertainty principle.
Note that, no matter whether $C$ in the commutation relation is an
operator or simply a constant, the right-hand side of the above
equation is always a constant. By applying this to $y$ and $l$ or
directly to $\eta$ and $\lambda$, we obtain
\begin{equation}
\Delta\eta\Delta\lambda\sim K,
\end{equation}
where $K$ is a constant (which is independent of the choice of
Lorentz-frames). As we shall see in Section IV, the uncertainty
principle and its immediate consequences indeed play an important
role in interpreting the obtained results.

\section{Probing stability, stationarity and scaling}
As far as the idea and the mathematical basis are concerned, both
Bachelier's Gaussian hypothesis \cite{3} and Mandelbrot's
hypothesis \cite{4,5,6} have to be considered as master-strokes,
yet what really counts is whether the relevant hypothesis indeed
describes the empirical data. In this connection, it has been
observed that the Gaussian hypothesis \cite{3} can, for example,
neither explain the typical erratic behavior of the empirical
sample second moments, nor reproduce the extremely heavy tails of
the empirical distributions. On the other hand, Mandelbrot's
fractal approach to Finance and Economics \cite{4,5,6} is well
supported by empirical facts. This conclusion is reached after a
series of detailed comparisons between empirical data and the
characteristic features of the Mandelbrot's theory, namely (A)
stability (B) stationarity and (C) scaling, has been performed
\cite{5,6,4}.

Taken together with the facts mentioned in Section III, these
comparisons have led us to the following question: Can the
concepts and methods introduced to Finance and Economics by
Bachelier and by Mandelbrot be helpful in studying and
understanding fluctuations in subnuclear reactions?

We recall that pseudorapidity $\eta$ (or rapidity $y$) is a
continuous variable, and that the multiplicity $dN/d\eta(\eta)$ of
charged hadrons at any $\eta$ (within the allowed kinematical
range $\eta_{min}\leq\eta\leq\eta_{max}$) can be very well
measured. Hence it is not only meaningful but also possible to
consider for a given event the multiplicity as well as its
relative changes in an interval $\Delta \eta$ of any size [for
details, see below especially item (d)] in the allowed kinematical
range. Based on the facts mentioned above, it seems tempting to
introduce, by analogy to Mandelbrot's $L_{M}(t,T)$ shown in Eq.
(2), the quantity:
\begin{equation}
 L(\eta,\Delta\eta)=\ln\frac{dN}{d\eta}(\eta+\Delta\eta)-\ln\frac{dN}{d\eta}(\eta),
\end{equation}
and to examine its properties. What we see, as far as the
kinematical limits and the experimental resolution allow, are the
following. (a) The $L(\eta,\Delta\eta)$'s can be {\it any} real
number. (b) The $L(\eta,\Delta\eta)$'s are summable, in the sense
that the adjacent $\eta$-interval can be combined or divided in an
arbitrary manner without having any influence on the definition of
$L(\eta,\Delta\eta)$ given in Eq. (11). (c) The set of
$L(\eta,\Delta\eta)$'s can be viewed as approximately
statistically independent, identically distributed random
variables. It follows then from the {\it Generalized} Central
Limit Theorem (cf. Appendix A and the references given therein)
that the only possible class of limiting distributions should be
{\it stable distributions}, which reduce to the special case
Gaussian if the $L(\eta,\Delta\eta)$'s have not only a mean, but
also a finite variance. (d) Since discontinuities are allowed to
occur in Mandelbrot's theory, the bin-size can be chosen to be so
small that even $dN/d\eta$=0 can be included. It should be pointed
out, however, that $dN/d\eta=0$ does not occur in the JACEE-data
as long as the bin-size is not chosen to be smaller than $\Delta
\eta=0.1$ which stands for the resolution power of these
experiments. (e)The variables $\eta$ and $\Delta\eta$ in
$L(\eta,\Delta\eta)$ are, in contrast to $t$ and $T\equiv\Delta t$
in Bachlier's $L_{B}(t,T)$ and Mandelbrot's $L_{M}(t,T)$, {\it
not} part of space-time, but as we have mentioned in Section III,
the dynamics of the system of produced hadrons should {\it not}
depend on whether we choose space-time or their canonical
conjugates as independent variables.

In the second part of this section, we perform a detailed
comparison between our theoretical expectations and the available
data, and show that the empirical distributions obtained from the
relevant experimental data are not only {\it stable} but also {\it
non-Gaussian}. Furthermore, we present evidence for {\it
stationarity} and {\it scale-invariance} ({\it scaling}).

We consider the cosmic-ray data obtained by JACEE-Collaboration
\cite{7} for Si+AgBr at 4 $TeV/nucleon$ and those for Ca+C (or O)
at 100 $TeV/nucleon$. These data have attracted much attention not
only because they show results of hadron-production processes in
nucleus-nucleus collisions obtained at energies much higher than
those obtained from accelerator-experiments, but also because they
exhibit unusually high multiplicities as well as significant
fluctuations in rapidity-distributions measured in more than 8 to
10 units of $\eta$. Among the theoretical discussions on these two
single-events \cite{7}, the work by Takagi \cite{13} and that by
Bialas and Peschanski \cite{14} are the most well-known ones. In
fact, their \cite{13,14} two-component conjecture for single-event
rapidity-distributions, in particular for those published by
JACEE-Collabortion \cite{7} has been adopted by most physicists
working in this field. According to their \cite{13,14} conjecture,
the observed rapidity-distribution of a given event can be, and
should be, viewed as a superposition of two distinct parts, namely
a ``statistical'' part and a ``nonstatistical'' or ``dynamical''
part. In Takagi's paper \cite{13}, the ``statistical'' part is
described ``by a smooth function with least oscillation'' where he
``assume(s) rather arbitrarily a function of the form $\cdots$''.
According to Bialas and Peschanski \cite{14}, the ``statistical''
part of such single-event distributions should be determined as
follows: ``If the inclusive rapidity spectrum (averaged over many
events) are not available as e.g. when one considers a single
high-multiplicity event, it is necessary to invoke one's
theoretical prejudices on the shape of the inclusive spectrum in
order to draw definite conclusions.''

In order to study the ``nonstatistical'' or ``dynamical'' part,
Takagi \cite{13} suggested that ``the rapidity density fluctuation
of nonstatistical origin may manifest itself as an oscillatory
pattern $\cdots$''. In his paper, the fluctuations are analyzed by
use of power spectrum. To study this part, Bialas and Peschanski
\cite{14} propose ``to study the dependence of factorial moments
of the rapidity distribution on the size of the resolution'' ``by
adapting the well-known results obtained in the investigation of
cascading phenomena and turbulent behavior''. Based on the
above-mentioned two-component conjecture \cite{13,14}, the random
cascading model \cite{14} and the method of factorial moments
\cite{14}, a large amount of experimental data has been analyzed,
and many refined models (including scaling models, multifractal
models etc.) have been developed. An overview of this development
together with the obtained results as well as detailed references
can be found in the review article by De Wolf, Dremin and Kittel
\cite{9}.

``Understanding'' fluctuations in empirical data by conjecturing
that the measured distribution of a given event consists of two or
more components (where every component fulfills its designed goal)
has also been rather popular in Finance and Economics, especially
in the pre-Mandelbrot era. Based on such a conjecture, many models
have been developed and very good fits to the data have been
produced. But, according to Mandelbrot \cite{5}: ``A common
feature of all these approaches, however, is that each new fact
necessitates an addition to the explanation. Since a new set of
parameters is thereby added, I don't doubt that reasonable
curve-fitting is achievable in many cases. '' In fact he clearly
expresses his doubts about the usefulness of such approach in
following terms \cite{4,5}. ``This form of symptomatic medicine (a
new drug for each complaint) could not be the last word!'' ``In my
view, even if an accumulation of quick fixes were to yield an
adequately fitting patchwork, it would bring no understanding.''

We begin our preconception-free data-analysis by noting that the
experimental resolution in the given histogram is 0.1, and by
choosing this to be the smallest $\eta$-interval, $\Delta \eta$.
In doing so, the $\eta$-range of the Si+AgBr event (hereafter
referred to as JACEE1) from $\eta=-4.0$ to 4.0 is divided into 80
$\Delta\eta$-bins, which lead to a set of 79
$L(\eta,\Delta\eta)$'s. Similarly, the $\eta$-range of the Ca+C
(or that of Ca+O) event (hereafter referred to as JACEE2) from
-5.0 to 5.0 is divided into 100 $\Delta\eta$-bins; hence there are
99 $L(\eta,\Delta\eta)$'s. In order to probe whether these
experimental values for $L(\eta,\Delta\eta)$'s can indeed be
considered as {\it stable} random variables which satisfy the {\it
Generalized} Central Limit Theorem, we make use of the definitions
and theorems summarized in Appendix A. To be more precise, we
check this in three steps.

In Step (i), we calculate the ``running sample mean'' of the $L(\eta,\Delta\eta)$'s
\begin{equation}
 \bar{L}_{n}(\Delta\eta)=\frac{1}{n}\sum\limits_{i=1}^{n}L(\eta_{i},\Delta\eta).
\end{equation}
Here, $n$ stands for the ordering number of the $\Delta\eta$-bins
where $\eta_{i}=\eta_{min}+(i-1)\Delta \eta$ and the ordering
begins from $\eta=\eta_{min}$ which is -4.0 for JACEE1, and -5.0
for JACEE2 respectively. The results are shown in Fig. 1 and Fig.
2 respectively. They seem to have the tendency of approaching zero
for sufficiently large $n$ in accordance with the Law of Large
Numbers (cf. Appendix A).

In Step (ii), we divide each one of the two
$L(\eta,\Delta\eta)$-sets (JACEE1 and JACEE2) into two distinct
groups according to the sign of every individual
$L(\eta,\Delta\eta)$, and plot separately, in accordance with
Definition A.2 given in Appendix A, the right- and the left-tail
distributions $P(L>L^{+})$ and $P(L<L^{-})$ respectively. The
result for JACEE1 is shown in Fig. 3, and that for JACEE2 is shown
in Fig. 4. The purpose of performing the plots are twofold: First,
we can see whether the distribution exhibits left-right symmetry
and thus determine the skewness parameter $\beta$ in case the
distribution is indeed stable (cf. Theorem A.3 in Appendix A).
Second, we can use them as the basis for a detailed
stability-test. See Figs. 5 and 6 for details.

In Step (iii), we start with the definition of stability
\begin{equation}
 S_{m}\overset{d}{=}c_{m}L+\gamma_{m}
\end{equation}
and note that in accordance with Definition A.1 and Theorem A.1,
the necessary and sufficient condition for the
$L(\eta,\Delta\eta)$'s to be {\it stable} random variables is that
for all integers $m>1$, there exist constants $c_{m}>0$ and
$\gamma_{m}$ such that Eq. (13) is true, where
\begin{eqnarray*}
S_{m}&=& \sum\limits_{i=1}^{m}L(\eta_{i},\Delta\eta) \\
&=& L(\eta,m\Delta\eta) \\
&=& ln\frac{dN}{d\eta}(\eta+m\Delta\eta)-ln\frac{dN}{d\eta}(\eta)
\end{eqnarray*}
and the $L(\eta_{i},\Delta\eta)$'s are identical copies of
$L(\eta,\Delta\eta)$ and $c_{m}=m^{1/\alpha}$ with $0<\alpha\leq
2$ and $\gamma_{m}$ is an arbitrary real number. In order to probe
the validity of Eq. (13), we choose the distributions mentioned in
``$\overset{d}{=}$'' to be tail distributions. As the first step,
we calculate and plot the right-hand-side of
\begin{equation}
c_{m}^{-1}(S_{m}-\gamma_{m})\overset{d}{=}L
\end{equation}
for the sets of $L(\eta,\Delta\eta)$'s obtained from JACEE1 and
that of JACEE2 respectively (they are of course the same as those
shown in Fig. 3 and Fig. 4. The purpose of reproducing them here
will become clear below). Next, we set $m$ on the left-hand-side
of Eq. (14) to be 2, 3, 4, 5 and 6 which corresponds to the
$\eta$-interval $m\Delta\eta$ respectively (that is $m$ times
0.1), and determine in every case the most suitable $c_{m}$ and
$\gamma_{m}$ (m=2, 3, 4, 5 and 6). We then use the method
described in Step (ii) to calculate the corresponding
$c_{m}^{-1}(S_{m}-\gamma_{m})^{+}$'s and
$c_{m}^{-1}(S_{m}-\gamma_{m})^{-}$'s and plot for every $m$ its
right-tail distribution
$P(c_{m}^{-1}[S_{m}-\gamma_{m}]>c_{m}^{-1}[S_{m}-\gamma_{m}]^{+})$,
and its left-tail distribution
$P(c_{m}^{-1}[S_{m}-\gamma_{m}]<c_{m}^{-1}[S_{m}-\gamma_{m}]^{-})$.
It turns out that in these two JACEE-events the maximum of $m$ is
$m=6$, because, beyond this, the sample size would not be large
enough to have reasonable statistics. Finally, we compare the
results obtained through the procedure mentioned above in the same
figure (here we see why we need the $m=1$ case in the same scale).
The results for JACEE1 are shown in Fig. 5, and those for JACEE2
are shown in Fig. 6. The striking agreement between the
distributions of the two sides of Eq. (14) shows that the
distributions of random variable $L(\eta,\Delta\eta)$'s under
consideration are {\it indeed} stable distributions.

It should be pointed out that, with the help of the mathematical
tools quoted in Appendix A, further information about these stable
distributions can be drawn from the results shown in Figs. 1-4: In
particular, in accordance with the Law of Large Numbers, Figs. 1
and 2 strongly suggest the existence of zero mean for JACEE1 and
JACEE2. Hence it seems meaningful to check, whether the
corresponding distributions are symmetric with respect to the mean
value and thus determine the skewness parameter $\beta$.  The fact
that the right-tail distributions and the left-tail distributions
are approximately equal as we can explicitly see in Figs. 3 and 4,
that is $P(L>L^{+})=P(L<L^{-})$ for JACEE1 and for JACEE2
respectively, shows that $\beta=0$ in both cases.

Next, we probe stationarity, that is, try to find out whether the
sets of $L(\eta,\Delta\eta)$'s obtained from the
$\eta$-distributions measured at different times (or
time-intervals) have {\it the same} statistical properties. Since
we do not have many sets of $\eta$-distribution data available,
what we {\it can} do {\it at present} is only to compare the set
obtained from JACEE1 with that obtained from JACEE2. In Fig. 7,
$P(L>L^{+})$ and $P(L<L^{-})$ for these two events are plotted
together in the same scale. What we see is that, both the right-
and left-tail distributions obtained from JACEE1 are very much the
same as those obtained from JACEE2---in agreement with the
theoretical expectations. The fact that these two events occurred
at different times and in reactions at different energies by using
different projectiles and targets makes the observed similarity
particularly striking!

Due to the role played by the concept of {\it scale-invariance}
({\it scaling}) in the fractal approach to Finance and Economics
and its role played in other Sciences including e.g. Condensed
Matter Physics, it is quite natural to ask: Are there also
empirical evidences, or at least indications, for {\it scaling} in
subnuclear reactions?

One way of checking this is to recall and to apply the method
proposed by Mandelbrot \cite{6} in analyzing the spot prices of
cotton. He performed such tests in two steps:

{\it In Step 1}, he considers the second moment of the daily
change of $\ln z(t)$ with $z(t)$ the spot price. Dividing the
period under consideration into 30 successive fifty-day samples
which can be numbered in chronological order $m$, he calculates
the second moment of every sample, and plot them against $m$ (cf.
Fig. 1 of Ref. [6]). Here, he sees enormous variability in time of
the sample second moment, and no tendency of any limiting
behavior.

{\it In Step 2}, he evaluates the cumulated absolute-frequency
distribution by making use of the above-mentioned figure and plot
the result on a double logarithmic paper (cf. Fig. 2 of Ref. [6])
in order to check whether the obtained points approximately lie on
one straight line. This is because, mathematically, any quantity
$N$ that can be expressed as some power of another quantity $s$,
$N(s)=s^{-\tau}$, has the following property. By taking the
logarithm on both sides, this equation yields $lnN(s)=-\tau lns$.
Hence the scale invariance can be seen from the simple fact that
the straight line looks the same everywhere. there are no features
at some scale which make that particular scale stand out; that is,
as $lns$ varies, $lnN$ shows no kinks or bumps anywhere. What he
sees (in Fig. 2 of Ref [6]) is indeed a straight line!

Following Mandelbrot's method, we divide the $\eta$-range in
JACEE1 and that in JACEE2 into successive 20-bin equal size
samples and give everyone of these samples an ordering number $m$.
Due to the fact that, the mean-values of large samples tend to
zero (see Fig. 1 and Fig. 2), the second moment (which corresponds
to that shown in Fig. 1 of Mandelbrot's paper \cite{6}) is
approximately equal to
\begin{equation}
S^{2}_{20}(\Delta
\eta)=\frac{1}{20}\sum\limits^{20}_{i=1}[L(\eta_{i},\Delta
\eta)-\bar{L}_{20}(\Delta \eta)]^{2}
\end{equation}
which stands for the variance in these two cases. We plot the
variance obtained from JACEE1 in Fig. 8, and that from JACEE2 in
Fig. 9. As we can explicitly see, their behavior are indeed rather
erratic. The corresponding frequency
distributions$$P(S^{2}_{20}>s^{2}_{20})$$ are plotted in Figs. 10
and 11 respectively. It seems that, in both cases the data
indicate the existence of power-law behavior and thus {\it
scale-invariance} ({\it scaling}). It should be pointed out,
however, since the number of data points in the JACEE-events are
much less than those for cotton-prices in Mandelbrot's analyses,
not only the sample-size in our case have to be smaller, but also
that the statistics are not as good as those for cotton-prices. In
order to amend this deficiency, we propose to use the following
alternative method.

Let us consider (instead of the sample variance given in
Fig. 1 of Ref. [6], and those used in Figs. 8 and 9 in this paper) the
``running sample variance''
\begin{equation}
   S_{n}^{2}(\Delta\eta)=\frac{1}{n-1}\sum\limits_{i=1}^{n}[L(\eta_{i},\Delta\eta)-\bar{L}_{n}(\Delta\eta)]^{2},
\end{equation}
where $\bar{L}_{n}(\Delta\eta)$ stands for the ``running sample
mean'' shown in Eq. (12) and Figs. 1 and 2. The range of the
ordering number $n$ is $n=2,3,\cdots,79$ for JACEE1, and
$n=2,3,\cdots,99$ for JACEE2. The ``running sample variance'' as a
function of the ordering number $n$ is shown in Fig. 12 and Fig.
13. Also in these figures, we see that the sequential sample
moment changes rather {\it erratically} with respect to $n$, and
does {\it not} seem to tend to any limit. That is to say,
erraticity and non-existence of tendency are nevertheless
characteristic properties of the plot, although running samples
are in general less independent than those obtained  by dividing
the original set into parts (as it is for example the case in
Fig.1 of Ref. [6] and in Figs. 8 and 9 of this paper). In order to
see explicitly that running samples do not have much influence on
the characteristic features of the original set, we also consider
in the JACEE1 case a set of 79 (and similarly in the JACEE2 case a
set of 99) Gaussian random variables with zero population mean and
unit population variance; and we examine in particular the
``running sample variance'' of the set of 79 (and the set of 99)
Gaussian random variables. What we see (not shown in this paper)
is of course distinct differences between the behavior of Gaussian
random variables and the that of JACEE-data \cite{7}. A more
effective way to examine the scaling behavior of such sets of
random variables is to plot the frequency distributions
$$P(S^{2}_{n}>s^{2}_{n})$$ on $\log-\log$ papers. Such plots for the
JACEE-data \cite{7} are shown in Fig. 14 and Fig. 15 respectively.
The 78 points obtained from JACEE1 and the 98 points obtained from
JACEE2 indeed lie approximately on straight lines in the
$\log-\log$ plots. Having the well-known relationship between
scaling, power-law behavior and straight-lines on double
logrithmatic papers in mind, the straight-line structure in the
$\log-\log$ plots and thus the property of {\it scale-invariance}
({\it scaling}) is evident. In order to see whether (and how much
if yes) the JACEE-data differ from Gaussian random variables, we
also use the above-mentioned sets of 79 and 99 Gaussian random
variables to do the same kind of plots, and show them in the
corresponding figure, that is in Fig. 14 and Fig. 15 respectively.
For such variables, the existence of a scale, namely
$\sigma^{2}=1$ (recall that we are considering Gaussian random
variables with zero population mean and unit population variance)
can be clearly seen [cf. Eqs. (A14) and (A15) in the appendix].

A number of conclusions can be drawn from the empirical results
plotted in Figs. 14 and 15, where once again use have been made of
the mathematical tools mentioned in Appendix A. First of all, the
result obtained from such $\log-\log$ plots has to be considered
as an effective indicator for scale-invariance (scaling): This is
because, in accordance with the Law of Large Numbers for sample
variance (see Theorem A.6 in the appendix), there is a profound
difference between sets of random variables with finite population
variance and those {\it without} finite population variance. While
in the former case, the running sample variance $S_{n}^{2}$ should
tend to the population variance $\sigma^{2}$ [see Eqs. (A14)] for
large $n$ which plays the role of a scale, there is {\it no} such
limiting behavior in the latter case. The empirical fact that the
frequency distributions of $S_{n}^{2}$ obtained from JACEE1 and
JACEE2 exhibit power-law behavior shows that the sets of
$L(\eta,\Delta\eta)$'s obtained from JACEE1 and JACEE2 {\it do not
have any given scale}. It can be readily seen from the
corresponding $\log-\log$ plots in which the data-points of each
JACEE-event lie on a straight line. This behavior is in sharp
contrast to that of a sample of Gaussian random variables of the
same size. Taken together with Definition A.4, A.5 and Theorem
A.4, A.6, the above-mentioned observation shows that the
population variance of the JACEE-sets should be infinite and that
the originally allowed range for the characteristic exponent
$\alpha$ namely $0<\alpha\leq 2$ should be narrowed down to
$0<\alpha<2$. In other words, the set of $L(\eta,\Delta\eta)$'s
obtained from JACEE1 and those obtained from JACEE2 are indeed
{\it non-Gaussian} stable random variables.

Furthermore, the results of stability tests shown in Figs. 5 and 6
can be, and should also be, considered as evidence for scale
invariance (scaling), because also they show that the relevant
scale---here the size of $\eta$-intervals---does not play a role.

In summary, what we have seen in the second part of this section
is that, the analysis of cosmic-ray data shows that the
$L(\eta,\Delta\eta)$'s defined in Eq. (11) which are the relative
changes of the multiplicities of charged hadrons between
successive rapidity-intervals $\Delta\eta$ can be considered as
mutually independent random variables, satisfying a {\it
non-Gaussian stable distribution which is stationary and
scale-invariant} ({\it scaling}). The striking similarity between
the properties of Mandelbrot's $L_{M}(t,T)$ and those of
$L(\eta,\Delta\eta)$ introduced in this paper suggests that the
underlying reaction mechanism(s) of the fluctuations in Financial
markets and those at the subnuclear level of matter are {\it very
much the same}. To be more specific, we note that everyone of the
$\Delta\eta$'s in $L(\eta,\Delta\eta)$ is related to its
corresponding $\Delta\lambda$ through Eq. (10), and the properties
of the $L(\eta,K/\Delta\lambda)$'s can be readily expressed in
terms of the $\Delta\lambda$'s. In particular, since
$\Delta\lambda$ decreases with increasing $\Delta\eta$,
measurements within larger and larger values of $\Delta\eta$
correspond to measurements within smaller and smaller values of
$\Delta\lambda$. It means, in this context, the validity of
stability, stationarity and scaling within larger and larger
$\Delta\eta$ intervals implies the validity of such
characteristics at smaller and smaller values of $\Delta\lambda$
(intervals in pseudolocality near the light-cone in space-time).
Having in mind that stability, stationarity and scaling are {\it
the} fundamental characteristics of Mandelbrot's fractal approach
to Finance, the result of the present empirical analysis should
perhaps be considered as an indication that {\it concepts and
methods used in Nonlinear Dynamics and/or in Complex Sciences
should be helpful in describing/understanding such hadronization
processes}. Studies along this line are now underway; the results
will be reported elsewhere.

\section{Concluding remarks}
The result obtained from analyzing the cosmic-ray data for
hadron-production, by using the concepts and methods introduced by
Bachelier and by Mandelbrot to Finance and Economics, shows that
it is not only possible but also useful to extract information on
subnuclear reactions directly from empirical distributions of
measurable quantities. It shows in particular that the
fluctuations in rapidity-distributions in the cosmic-ray data have
much in common with the fluctuations observed in stock market. The
striking similarity seems to suggest that they are complex
phenomena of {\it the same} ({\it fractal}) nature. The fact that
{\it non-Gaussian stable distributions which are stationary and
scale-invariant} ({\it scaling}) describe the existing data
remarkably well calls for further attention. It would be very
helpful to have comparison with data taken at other energies
and/or for other collision processes.

\begin{acknowledgments}
 The authors thank Cai Xu, Feng Youceng, Hwa Rudolf, Kittel Wolfram,
 Li Wei, Liang Zuotang, Peng Hong'an, Qian Wanyan, Ratti Sergio P., Sa Benhao,
 Schertzer Daniel, Yang Chunbin and Zhu Wei for helpful discussions and valuable comments.
 They  also thank National Natural Science Foundation of China and
 KeYanChu of CCNU for financial suport.
\end{acknowledgments}

\appendix

\section{}
This is a mathematical appendix added to the present paper to make
it self-contained. It contains a brief introduction to {\it stable
distributions} and a set of definitions and theorems (without
proofs) taken from monographs and/or textbooks related to this
subject \cite{15}.

The term ``stable distributions'' stands for a rich class of probability
distributions which allow skewness and heavy tails, and have a number of
other intriguing properties. One of them is
the lack of closed formulas for densities and distributions for all
but a few; such exceptional stable distributions are Gaussian,
Cauchy and L\'{e}vy distributions.


{\bf Definition A.1} Let $X_{1}, X_{2},\cdots, X_{n}$ and $X$ be mutually
independent random variables with a common distribution $P$ which is not
concentrated at one point, where $n>1$ and $S_{n}$ is the sum of these $n$
random variables $S_{n}=X_{1}+X_{2} +\cdots+X{n}$.
Then, the distribution $P$ is {\it stable in the broad sense}
if and only if for each $n$, there exist constants $c_{n}>0$ and $\gamma_{n}$ such that
\begin{equation}
 S_{n}\overset{d}{=}c_{n}X+\gamma_{n}.
\end{equation}
Here, the symbol ``$\overset{d}{=}$'' means equality in
distribution, i.e. both expresses obey the same probability law.
The distribution $P$ is {\it stable in the strict sense} if and
only if $\gamma_{n}=0$ for all $n$.

{\bf Theorem A.1} The norming constants $c_{n}$ are of the form $c_{n}=n^{1/\alpha}$
with $0<\alpha\leq 2$, where the constant $\alpha$ is called the
characteristic exponent of $P$.

{\bf Theorem A.2} A random variable $X$ is stable if and only if $X \overset{d}{=}AZ+B$,
where $0<\alpha\leq 2$, $-1\leq\beta\leq 1$, $A>0$, $B\in\Re$ and $Z$
is a random variable with characteristic function 
\begin{equation}
E\exp(iuZ)=\exp(-\left|u\right|^{\alpha}[1-i\beta\tan\frac{\pi\alpha}{2}(sign
u)])\quad {\alpha}\neq1
\end{equation}
\begin{equation}
E\exp(iuZ)=\exp(-\left|u\right|[1+i\beta\frac{2}{\pi}(sign
u)\ln\left|u\right|])\quad {\alpha}=1.
\end{equation}


{\bf Theorem A.3} A general stable distribution requires four parameters to describe.
The commonly used set is $\{\alpha,\beta,\gamma,\delta\}$. Here $0<\alpha\leq 2$ is known as
the index of stability or the characteristic exponent; $-1\leq\beta\leq 1$ the skewness
parameter; $\gamma>0$ the scale parameter; and $\delta\in\Re$ the location parameter. The
three exceptional distributions mentioned at the beginning of the appendix, namely \\
(a) Normal or Gaussian distribution, $X\sim N(\mu,\sigma^{2})$ with a density
\begin{equation}
 f(x)=\frac{1}{\sqrt{2\pi}\sigma}\exp(-\frac{(x-\mu)^{2}}{2\sigma^{2}}),\quad
 -\infty<x<\infty,
\end{equation}
(b) Cauchy distribution, $X\sim Cauchy(\gamma,\delta)$ with a density
\begin{equation}
 f(x)=\frac{1}{\pi}\frac{\gamma}{\gamma^{2}+(x-\delta)^{2}}, \quad
 -\infty<x<\infty,
\end{equation}
(c) L\'{e}vy distribution, $X\sim L$\'{e}$vy(\gamma,\delta)$ with a density
\begin{equation}
 f(x)=\sqrt{\frac{\gamma}{2\pi}}\frac{1}{(x-\delta)^{3/2}}\exp(-\frac{\gamma}{2(x-\delta)}),\quad
 \delta<x<\infty,
\end{equation}
are special cases of stable distributions. Also their
characteristic functions have precisely the form given in Eqs.
(A2) or (A3) mentioned in Theorem A.2. The parameters have
respectively the following set of values:
\begin{eqnarray*}
 & X & \sim N(\mu,\sigma^{2}): \alpha=2, \beta=0, A=\sigma^{2}/2, B=\mu; \\
 & X & \sim Cauchy(\gamma,\delta): \alpha=1, \beta=0, A=\gamma, B=\delta; \\
 & X & \sim \text{L\'{e}vy}(\gamma,\delta): \alpha=1/2, \beta=1, A=\gamma, B=\delta.
\end{eqnarray*}

{\bf Definition A.2} Let $X$ be a random variable with distribution $P(x)$.
Then its {\it right-tail distribution} $T_{r}(x)$ is
\begin{equation}
  T_{r}(x)=1-P(x)=P(X>x)
\end{equation}
where $x\in\Re$ is any real number. The corresponding {\it left-tail distribution} $T_{l}(x)$ is
\begin{equation}
  T_{l}(x)=P(x)=P(X<x).
\end{equation}

{\bf Definition A.3} A distribution is said to be {\it heavy tailed}
if its tails are heavier than exponential. For $\alpha<2$, stable distributions
have one tail (when $\alpha<1$ and $\beta=\pm 1$) or both tails (all other cases)
that are asymptotically power laws with heavy tails.

{\bf Definition A.4} Let $f(x)$ be the density function of a stable distribution.
The expression
\begin{equation}
E(\left|X\right|^{p})=\int\limits_{-\infty}^{\infty}dx\left|x\right|^{p}f(x)
\end{equation}
is called the fractional absolute moment of this distribution, where $p$ can be
any real number.

{\bf Theorem A.4} The fractional absolute moment
$E(\left|X\right|^{p})$ is finite if and only if $0<p<\alpha$
where $\alpha$ is the characteristic exponent of this stable
distribution (which satisfies $0<\alpha<2$), and that
$E(\left|X\right|)^{p}$ is {\it infinite} for $p\geq\alpha$.

{\bf Theorem A.5} Khinchine's Law of Large Numbers

Let $\{X_{i} (i=1,2,\cdots,n)\}$ be a sequence of mutually
independent random variables with a common distribution. If the
expectation $\mu=E(X_{i})$ exists, then for every $\varepsilon
>0$, the probability
\begin{equation}
  P\{\left|\frac{X_{1}+X_{2}+\cdots+X_{n}}{n}-\mu\right|>\varepsilon\}\rightarrow0 \quad for \quad n\rightarrow\infty.
\end{equation}
It means in the language of Mathematical Statistics: As $n$ tends to infinity,
the expression $\frac{1}{n}\sum\limits_{i=1}^{n}X_{i}$ converges in probability
to $\mu$; and this is often written as:
\begin{equation}
  \frac{1}{n}\sum\limits_{i=1}^{n}X_{i}\overset{P}{\longrightarrow}\mu \quad for \quad n\rightarrow \infty.
\end{equation}

{\bf Definition A.5} Let $\{X_{i} (i=1,2,\cdots,n)\}$ be a sample of size $n$ taken
from the population of a random variable $X$ with population mean $E(X)=\mu<\infty$
and population variance $D(X)=\sigma^{2}<\infty$. Then, the expression
\begin{equation}
 \bar{X}_{n}=\frac{1}{n}\sum\limits_{i=1}^{n}X_{i},
\end{equation}
which is nothing else but the expression on the left-hand-side of the arrow
in (A11), and
\begin{equation}
 S_{n}^{2}=\frac{1}{n-1}\sum\limits_{i=1}^{n}(X_{i}-\bar{X}_{n})^{2}
\end{equation}
are called, respectively, the ``running sample mean''
and the ``running sample variance'' of $X$.

{\bf Theorem A.6} Law of Large Numbers for sample variance

The sample variance $S_{n}^{2}$ defined in Eq. (A13) with $\bar{X}_{n}$ given
by (A12) behaves, for increasing $n$, as follows:
\begin{equation}
  S_{n}^{2}\overset{P}{\longrightarrow} \sigma^{2} \quad for \quad n\rightarrow
  \infty,
\end{equation}
and for all intergers $n$, we have:
\begin{equation}
  E(S_{n}^{2})=\sigma^{2}.
\end{equation}

{\bf Theorem A.7} Central Limit Theorem

Let $\{X_{i} (i=1,2,\cdots,n)\}$ be a sequence of mutually independent random variables
with a common distribution. If the expectation $\mu=E(X_{i})$ and the
variance $\sigma^{2}=Var(X_{i})$ exist, then the sum $S_{n}=X_{1}+X_{2}+\cdots+X_{n}$
has the following limiting behavior. For every fixed $x$,
\begin{equation}
 \lim_{n\rightarrow\infty}P\{\frac{S_{n}-n\mu}{\sigma\sqrt{n}}<x\}=
 \frac{1}{\sqrt{2\pi}}\int\limits_{-\infty}^{x}dt\exp(-\frac{t^{2}}{2}).
\end{equation}

This theorem is a special case of the following theorem.

{\bf Theorem A.8} Generalized Central Limit Theorem

Let $\{X_{i} (i=1,2,\cdots,n)\}$ be a sequence of mutually independent,
identically distributed random variables. Then, there exist constants
$a_{n}>0$, $b_{n}\in\Re$ and a non-degenerated random variable $Z$ with
\begin{equation}
 a_{n}(X_{1}+X_{2}+\cdots+X_{n})-b_{n}\overset{d}{\longrightarrow} Z
\end{equation}
if and only if $Z$ is stable; in which case $a_{n}=n^{-1/\alpha}$ for $0<\alpha\leq 2$.

\newpage 
\bibliography{apssamp}

\newpage
\begin{figure}[tbph]
\includegraphics{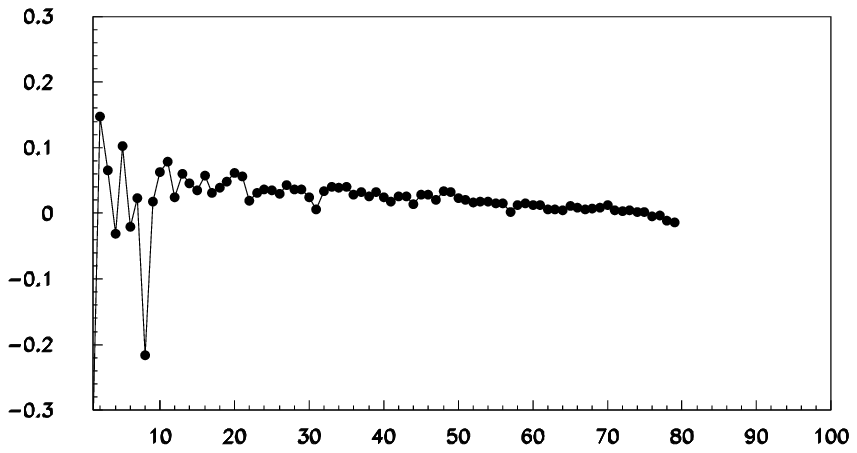}
\caption{Running sample mean $\bar{L}_{n}(\Delta\eta)$ of the
$L(\eta,\Delta\eta)$'s [see Eq. (12)] for $\Delta\eta=0.1$ (the
experimental resolution) is plotted as function of the ordering
number $n$. Data are taken from JACEE1 \cite{7}. Here,
$\eta_{min}=-4.0$ is the starting point for the ordering of the 80
$\Delta\eta$-bins. The solid line indicates the ideal case in
which $\bar{L}_{n}\equiv 0$.} \label{fig1}
\end{figure}

\begin{figure}[tbph]
\includegraphics{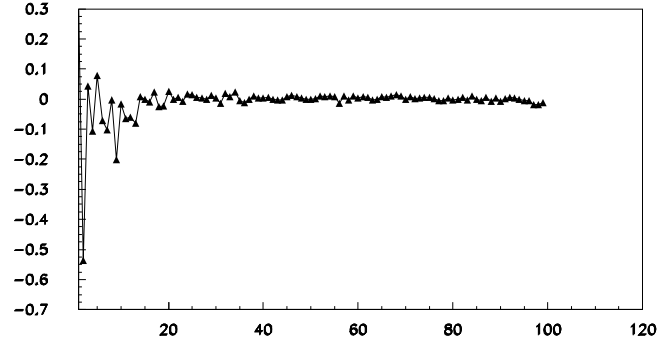}
\caption{Running sample mean $\bar{L}_{n}(\Delta\eta)$ of the
$L(\eta,\Delta\eta)$'s for $\Delta\eta=0.1$ (the experimental
resolution) is plotted as function of $n$. Data are taken from
JACEE2 \cite{7}. Here, $\eta_{min}=-5.0$ and there are 100
$\Delta\eta$-bins. The solid line indicates the ideal case in
which $\bar{L}_{n}\equiv 0$.} \label{fig2}
\end{figure}

\begin{figure}[tbph]
\includegraphics{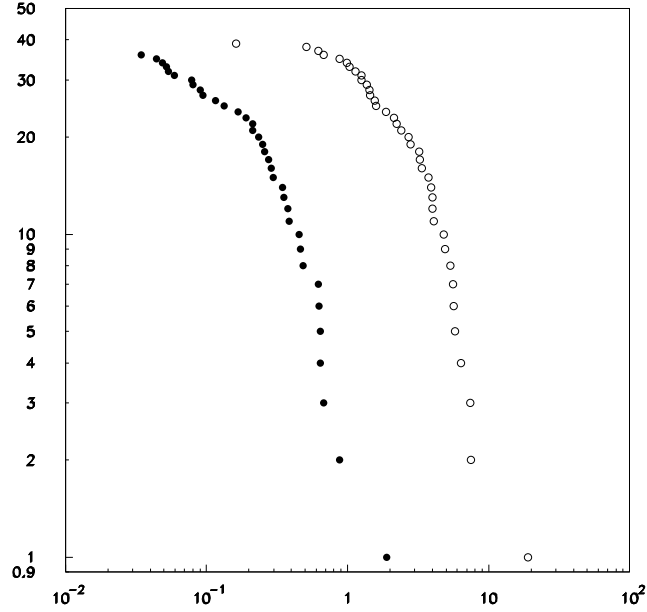}
\caption{Tail-distributions of the $L(\eta,\Delta\eta)$'s for
JACEE1 \cite{7}: The right-tail distribution $P(L>L^{+})$ is shown
as a set of black dots. $P(L>\left|L^{-}\right|)$ which is the
mirror image of the left-tail distribution $P(L<L^{-})$ is shown
as a set of open circles. They are plotted as functions of $L^{+}$
and $\left|L^{-}\right|$ respectively. Here,
$P(L>\left|L^{-}\right|)$ is plotted instead of $P(L<L^{-})$ in
order to avoid the problem of taking logarithm of negative values;
the values of $\left|L^{-}\right|$ are multiplied by a factor 10
so that the two sets of points can be kept away from one another
in the figure. This technique used in plotting $P(L<L^{-})$ and
comparing it with the corresponding $P(L>L^{+})$ in the same
figure is used throughout the paper whenever it is needed.}
\label{fig3}
\end{figure}

\begin{figure}
\includegraphics{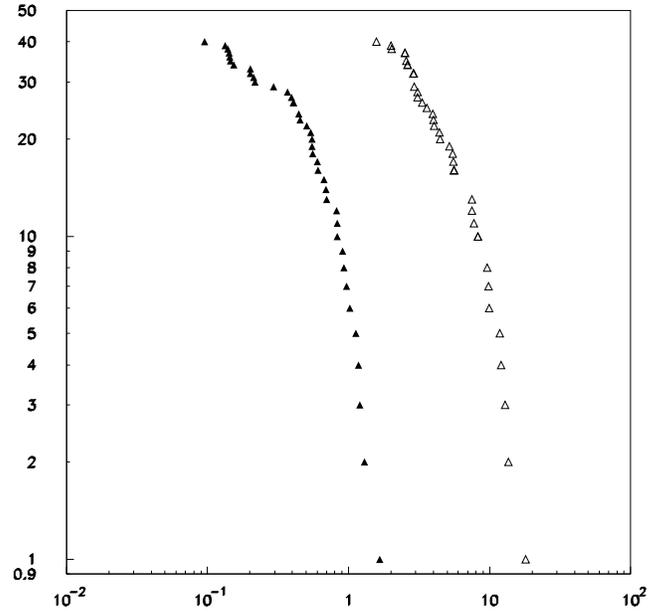}
\caption{Tail-distributions of the $L(\eta,\Delta\eta)$'s for
JACEE2 \cite{7}: Here the right-tail distribution is shown as a
set of black triangles; while those for the left-tail are shown as
open triangles. The technique used in plotting $P(L<L^{-})$ used
here are the same as that used in Fig. 3.} \label{fig4}
\end{figure}

\begin{figure}[tbph]
\includegraphics[width=0.6\textwidth]{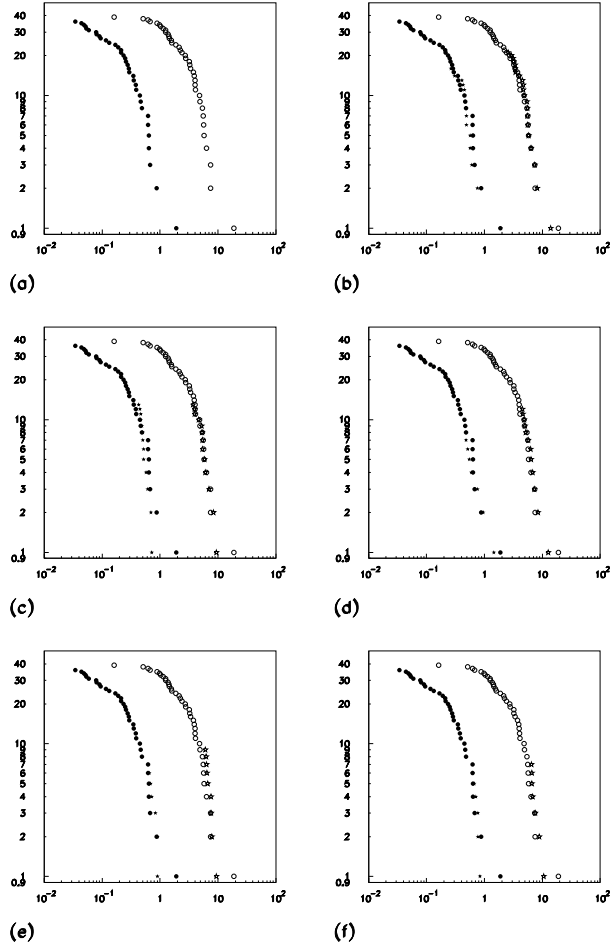}
\caption{Direct stability tests for JACEE1 \cite{7}:  Figs. 5(b)
to 5(f) correspond respectively to the $m=2,3,\cdots,6$ cases
mentioned in Section IV Step (iii). Here, $m$ stands for the
number in equal-size-divided samples in the set of
$L(\eta,\Delta\eta)$'s. This is obviously the same $m$ as that in
Eq. (14) which expresses the necessary and sufficient condition of
stability for the $L(\eta,\Delta\eta)$'s under consideration. The
black dots and open circles in all these figures are respectively
the right-tail (r-t) and left-tail (l-t) distributions
$P(L>L^{+})$ and $P(L>\left|L^{-}\right|)$ as functions of $L^{+}$
and $\left|L^{-}\right|$. In the figures 5(a) through 5(f) we
write simply r-t or l-t as functions of $L^{+}$ and
$\left|L^{-}\right|$ in order to safe space. They are to be
compared separately with the black stars and the open stars which
stand respectively for
$P(c_{m}^{-1}[S_{m}-\gamma_{m}]>c_{m}^{-1}[S_{m}-\gamma_{m}]^{+})$
and
$P(c_{m}^{-1}[S_{m}-\gamma_{m}]>\left|c_{m}^{-1}[S_{m}-\gamma_{m}]^{-}\right|)$
for $m=2,3,\cdots,6$. In the figures we write simply r-t or l-t as
functions of $N^{+}_{m}\equiv c_{m}^{-1}[S_{m}-\gamma_{m}]^{+}$ or
$\left|N^{-}\right|\equiv
\left|c_{m}^{-1}[S_{m}-\gamma_{m}]^{-}\right|$ to safe space. Note
that the quality of stability can be judged by the degree how good
the black stars/open stars overlap with the corresponding black
dots/open circles. Fig. 5(a) is a reproduction of Fig. 3 in the
same scale as Figs. 5(b) to 5(f), so that comparison in globe view
can be made.} \label{fig5}
\end{figure}

\begin{figure}[tbph]
\includegraphics[width=0.6\textwidth]{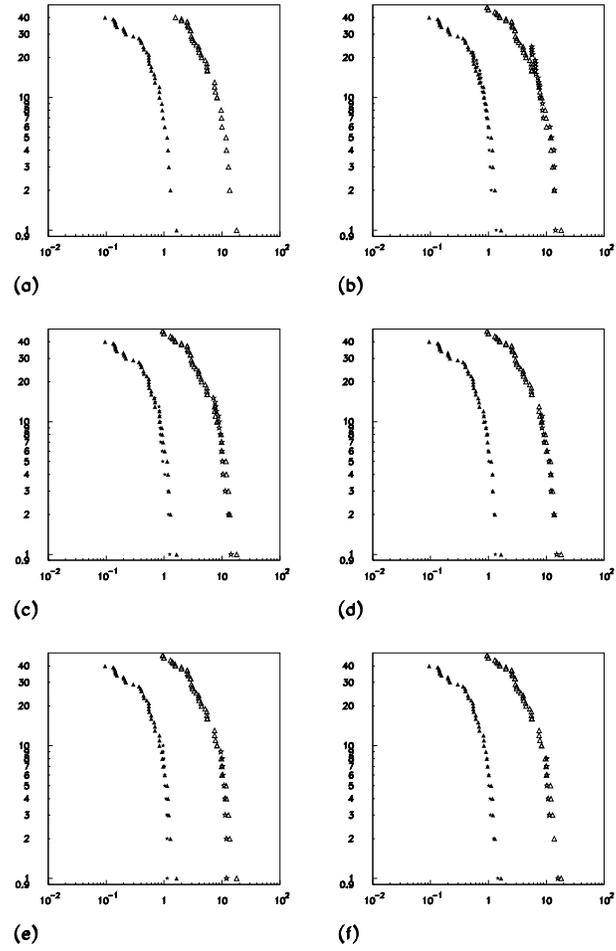}
\caption{Direct stability tests for JACEE2 \cite{7}: The notations
and the technique in plotting the tail-distributions used here are
the same as those used in Fig. 5, except that the black and open
circles should now be replaced by black and open triangles.}
\label{fig6}
\end{figure}

\begin{figure}[tbph]
\includegraphics{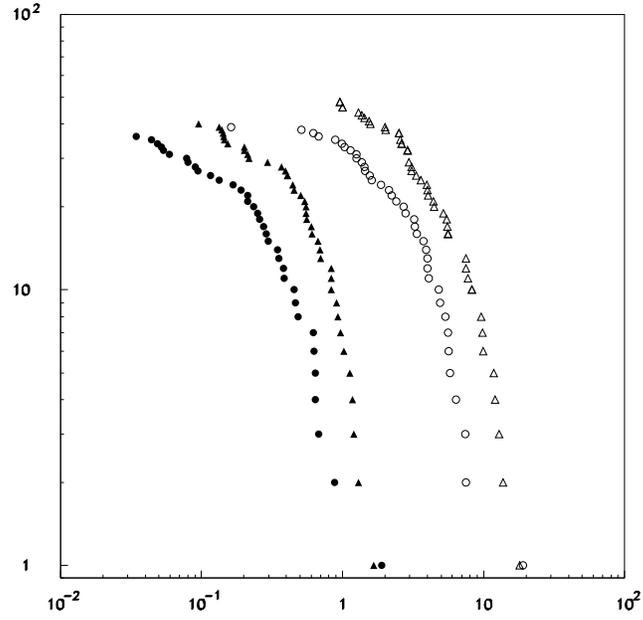}
\caption{Stationarity test: The right-tail (r-t) distribution
$P(L>L^{+})$  and the left-tail (l-t) distribution
$P(L>\left|L^{-}\right|)$ for JACEE1 (abbreviated as J1 in the
figure) \cite{7} are plotted in the same figure with those for
JACEE2 (abbreviated as J2 in the figure) \cite{7}. The JACEE1 data
are shown as black dots and open circles respectively, while the
corresponding JACEE2 data are shown as black and open triangles.}
\label{fig7}
\end{figure}

\begin{figure}[tbph]
\includegraphics{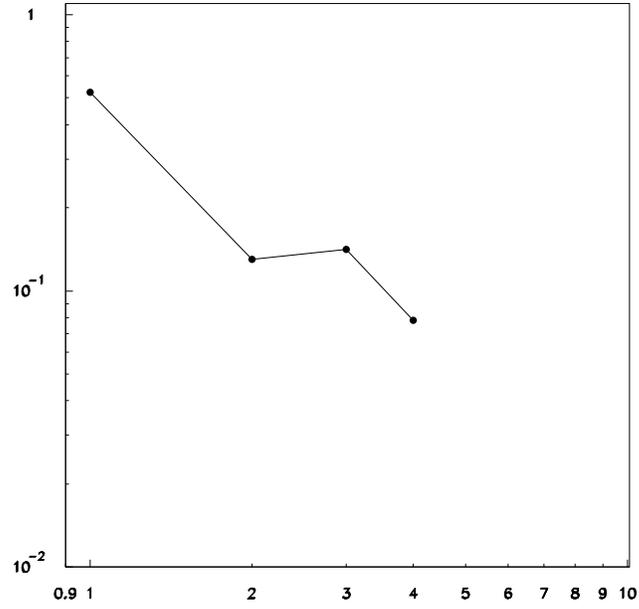}
\caption{The variance of samples of the same size (20),
$S_{20}^{2}(\Delta\eta=0.1)$, is plotted as function of the
ordering number $m$ for the $L(\eta,\Delta\eta)$'s obtained from
JACEE1 \cite{7}. Due to the limited number of data, $m$=1, 2, 3 or
4. This figure is the analogue of Fig. 1 of Ref. [6].}
\label{fig8}
\end{figure}

\begin{figure}[tbph]
\includegraphics{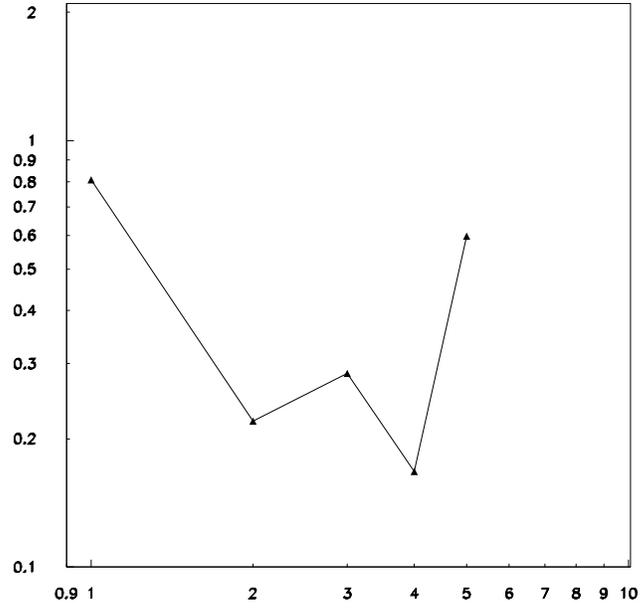}
\caption{$S_{20}^{2}(\Delta\eta=0.1)$ is plotted against $m$. Data
are taken from JACEE2 \cite{7}. Here, $m$=1, 2, 3, 4 or 5. Also
this figure is the analogue of Fig. 1 of Ref. [6].} \label{fig9}
\end{figure}

\begin{figure}[tbph]
\includegraphics{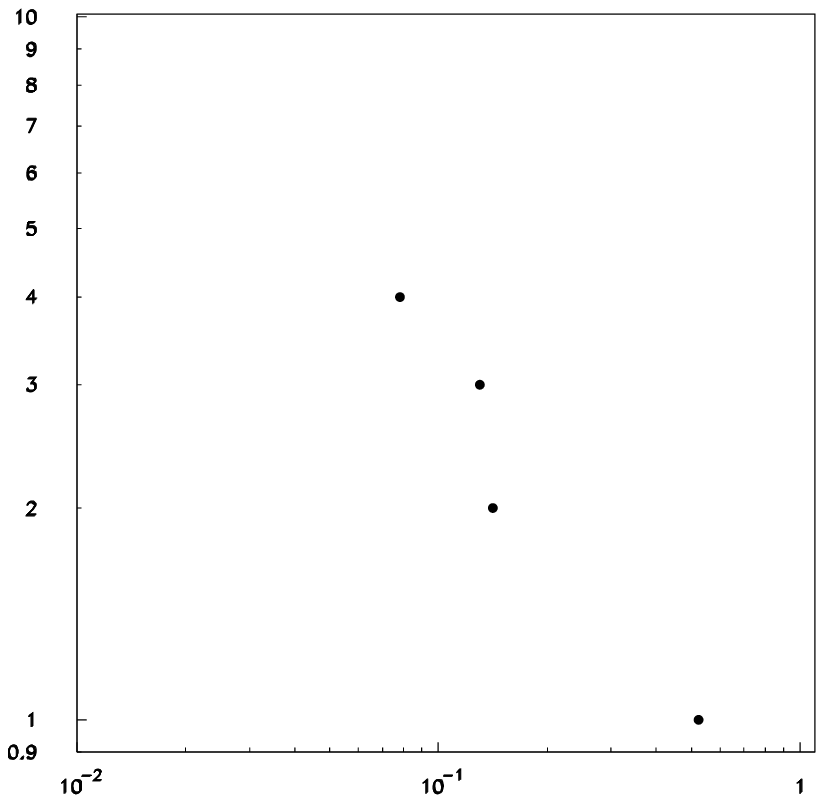}
\caption{Scaling test for JACEE1 \cite{7} by using the method
proposed by Mandelbrot \cite{6}: This is the cumulated absolute
frequency distribution for the $S_{20}^{2}$'s shown in Fig. 8.
This figure is the analogue of Fig. 2 of Ref. [6].} \label{fig10}
\end{figure}

\begin{figure}[tbph]
\includegraphics{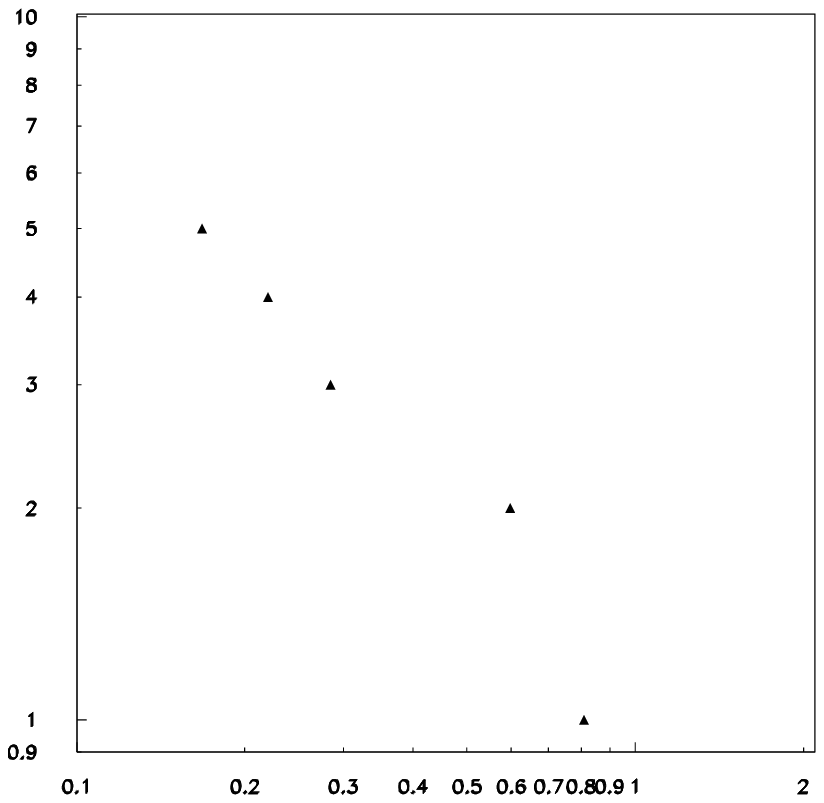}
\caption{Scaling test for JACEE2 \cite{7} by using the method
proposed by Mandelbrot \cite{6}: This is the cumulated absolute
frequency distribution for the $S_{20}^{2}$'s shown in Fig. 9.
Also this is the analogue of Fig. 2 of Ref. [6].} \label{fig11}
\end{figure}

\begin{figure}[tbph]
\includegraphics{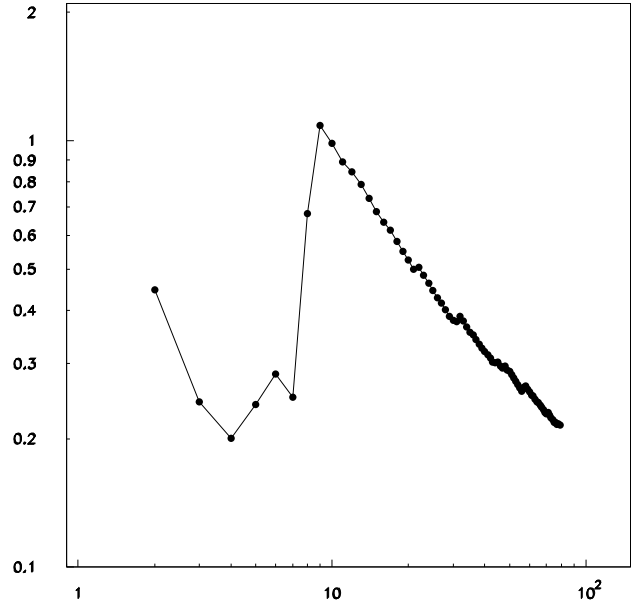}
\caption{The running sample variance $S_{n}^{2}(\Delta\eta=0.1)$
for the $L(\eta,\Delta\eta)$'s obtained from JACEE1 \cite{7} is
plotted as function of the ordering number $n$, where $n=2, 3,
\cdots, 79$.} \label{fig12}
\end{figure}

\begin{figure}[tbph]
\includegraphics{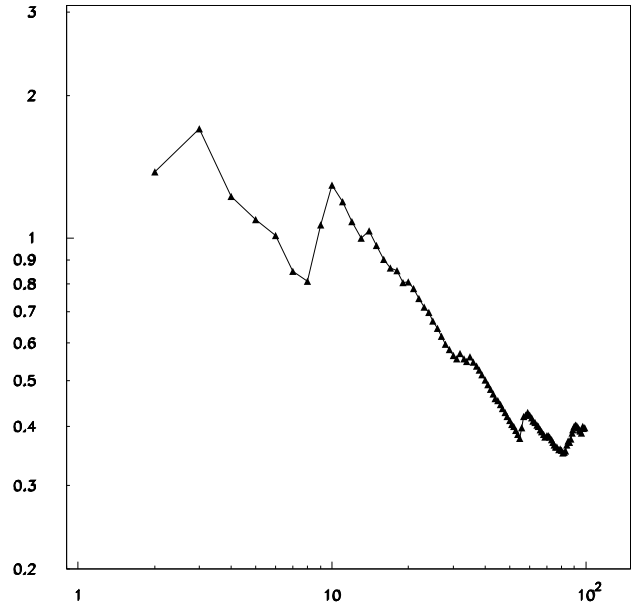}
\caption{The running sample variance $S_{n}^{2}(\Delta\eta=0.1)$
is plotted against $n$. Data are taken from JACEE2 \cite{7}, where
$n=2, 3, \cdots, 99$.} \label{fig13}
\end{figure}

\begin{figure}[tbph]
\includegraphics{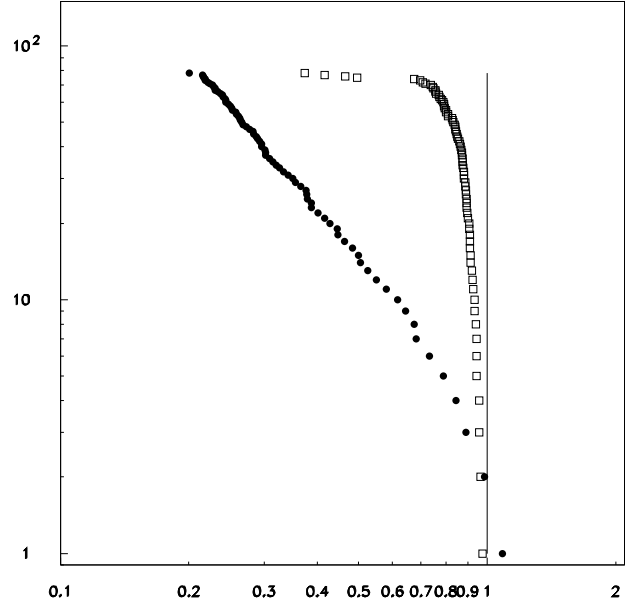}
\caption{Scaling test for JACEE1 \cite{7} by using the method
proposed in the present paper: The cumulated absolute frequency
distribution for the $S_{n}^{2}$'s (calculated and plotted in Fig.
12) is shown as black dots in this figure. For the sake of
comparison, a sample of the same size (79 for JACEE1) of standard
Gaussian random variables (with zero population mean and unity
population variance, $\sigma^{2}=1$) is also taken into account,
where the running sample variance $S_{n}^{2}$'s are also
calculated, and the corresponding cumulated absolute frequency
distribution (denoted by open squares) is plotted. The solid line
stands for the mean of $S_{n}^{2}$ of the sample of standard
Gaussian random variables for all $n$'s, that is for
$E(S_{n}^{2})=\sigma^{2}=1$. It also stands for the limiting value
of the sample variance, which is according to the Law of Large
Numbers: $S_{n}^{2}\overset{P}{\longrightarrow} \sigma^{2}=1$.}
\label{fig14}
\end{figure}

\begin{figure}[tbph]
\includegraphics{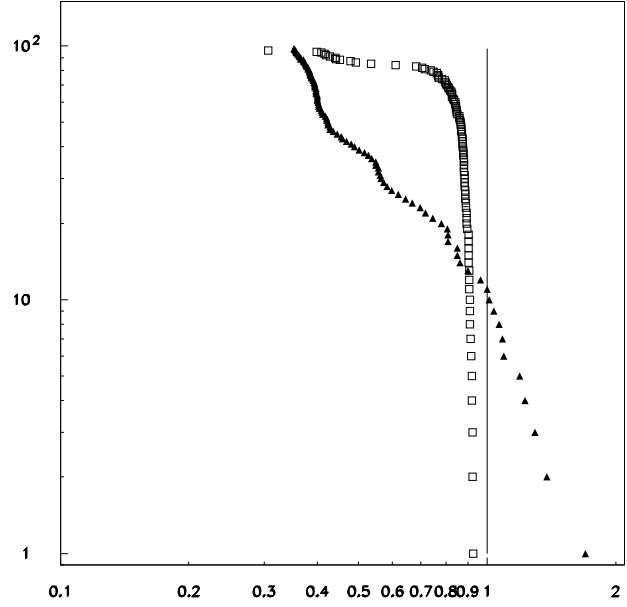}
\caption{Scaling test for JACEE2 \cite{7} by using the method
proposed in the present paper: The cumulated absolute frequency
for the $S_{n}^{2}$'s (calculated and plotted in Fig. 13) is shown
as black triangles in this figure. The notations used here are the
same as those used in Fig. 14. The size of the JACEE2 sample and
that of the corresponding Gaussian sample is 99.} \label{fig15}
\end{figure}

\end{document}